\documentstyle[epsfig,color,12pt]{article}

\include{epsf}

\def\beq{\begin{equation}}
\def\eeq{\end{equation}}

\begin{document}

\centerline{\large\bf Regular rotating electrically charged black
holes and solitons}

\centerline{\large\bf in nonlinear electrodynamics minimally
coupled to gravity}

 \vskip 0.1in

\centerline{ Irina~Dymnikova$^{a,b}$ and Evgeny Galaktionov$^{a}$}

\vskip 0.2in

\centerline{\it $^{a}$\sl A.F. Ioffe Physico-Technical Institute,
Politekhnicheskaja 26, St.Petersburg, 194021 Russia}

\centerline{\it $^{b}$\sl Department of Mathematics and Computer
Science, University of Warmia and Mazury,}

\centerline{\it S{\l}oneczna 54, 10-710  Olsztyn, Poland; e-mail:
irina@uwm.edu.pl}

\vskip0.1in

{\bf Abstract}

In nonlinear electrodynamics coupled to gravity, regular
spherically symmetric electrically charged solutions satisfy the
weak energy condition and have obligatory de Sitter centre.
 By the G\"urses-G\"ursey algorithm they are transformed to
spinning electrically charged solutions asymptotically Kerr-Newman
for a distant observer. Rotation transforms de Sitter center into
de Sitter vacuum surface which contains equatorial disk $r=0$ as a
bridge. We present general analysis of the horizons, ergoregions
and de Sitter surfaces, as well as the conditions of the existence
of regular solutions to the field equations. We find asymptotic
solutions and show that de Sitter vacuum surfaces have properties
of a perfect conductor and ideal diamagnetic, violation of the
weak energy condition is prevented by the basic requirement of
electrodynamics of continued media, and the Kerr ring singularity
is replaced with the superconducting current.

{\bf Journal Reference: Class. Quant. Grav. 32 (2015) 165015}

\vskip0.2in

\section{Introduction}

The Kerr-Newman electrically charged rotating solution to the
Maxwell-Einstein equations \cite{newman}
 $$
ds^2 = \frac{(2mr-e^2) - \Sigma}{\Sigma} dt^2  +
\frac{\Sigma}{\Delta} dr^2 + \Sigma d\theta^2 -
\frac{2a(2mr-e^2)\sin^2\theta}{\Sigma}dt d\phi
 $$
 \beq
 + \biggl(r^2 + a^2
+ \frac{(2mr-e^2)a^2\sin^2\theta}{\Sigma}\biggr)\sin^2\theta
d\phi^2; ~~ \Delta = r^2 - 2mr + a^2 + e^2 ,
                                                                              \label{kn}
  \eeq
where $\Sigma$-function and the associated electromagnetic
potential read
  \beq
 \Sigma = r^2 + a^2\cos^2\theta; ~~
 A_i = - \frac{e r}{\Sigma}[1; 0, 0, -a\sin^2\theta] ,
                                                                             \label{kn1}
 \eeq
was obtained from the Reissner-Nordstr\"om electrovacuum solution
with using the complex coordinate transformation discovered by
Newman and Janis \cite{trick}.

The Kerr-Newman geometry has at most two horizons
$r_{\pm}=m\pm\sqrt{m^2-(a^2+e^2)}$, and
 implies the same value of the gyromagnetic
 ratio $g = 2$, as predicted for a spinning particle by the Dirac equation
\cite{carter}. In the case appropriate for particles,
$a^2+e^2>m^2$,  there are no Killing horizons, the manifold is
geodesically complete (except for geodesics which reach the
singularity), and any point can be connected to any other point by
both a future and a past directed time-like curve. Closed
time-like curves originate in the region where $g_{\phi\phi} <0$,
can extend over the whole manifold and cannot be removed by taking
a covering space \cite{carter}.

The Kerr-Newman solution belongs to the Kerr family of solutions
to the source-free Maxwell-Einstein equations, and
 represents the exterior fields of a rotating charged body \cite{carter}.
 The source models for the Kerr-Newman exterior can be roughly
divided into disk-like \cite{werner,bur74,lopez1},
shell-like \cite{delacruz,cohen,lopez},
bag-like \cite{boyer,trumper,tiomno,bur89,bur2000,behm,bur2014},
and string-like (\cite{bur2013} and references therein).

The problem of matching the Kerr-Newman exterior to a rotating
material source does not have a unique solution, since one is free
 to choose arbitrarily the boundary between the exterior
and the interior \cite{werner}, not speaking about freedom in
choosing an interior model itself.

On the other hand, one can study equations describing a dynamical
system to get information about its basic properties rather than
postulate properties of a source which could give rise to the
Kerr-Newman fields. In the case of an electromagnetically
interacting structure the appropriate equations come from
nonlinear electrodynamics coupled to gravity (NED-GR)\footnote{NED
theories appear as low-energy effective limits in certain models
of string/M-theories \cite{fradkin,tseytlin,witten}.}.

Nonlinear electrodynamics was proposed by Born and Infeld  as
founded on two basic points: to consider electromagnetic field and
particles within the frame of one physical entity which is
electromagnetic field; to avoid letting physical quantities become
infinite \cite{born}. This program can be realized in nonlinear
electrodynamics coupled to gravity. Source-free NED-GR equations
admit regular causally safe axially symmetric asymptotically
Kerr-Newman solutions \cite{me2006}, which describe regular rotating charged
black holes and electromagnetic spinning solitons.

The key point is that for any gauge-invariant Lagrangian ${\cal
L}(F)$, stress-energy tensor of  electromagnetic field
  in
  the spherically symmetric case
has the algebraic structure
 \beq
T^t_t=T^r_r ~ ~ ~(p_r=-\rho) .
                                                                                \label{myvac}
  \eeq
Regular spherically symmetric solutions with stress-energy tensors
specified by (\ref{myvac}) satisfying the weak energy condition
(non-negativity of density as measured by any local observe), have
obligatory de Sitter center with $p=-\rho$
\cite{me92,me2000,me2002,me2003}.  The mass $m$ of an object is
generically related to de Sitter vacuum and breaking of space-time
symmetry from the de Sitter group in the origin \cite{me2002}.In
the NED-GR regular solutions interior de Sitter vacuum  provides  a
proper cut-off on self-interaction divergent for a point charge
\cite{me2004,me2006}.

The regular spherical solutions generated by (\ref{myvac}) belong
to the Kerr-Schild class \cite{ks,behm,ssqv} and can be
transformed by the G\"urses-G\"ursey algorithm \cite{gurses} into
regular axially symmetric solutions which describe regular
rotating electrically charged objects, asymptotically Kerr-Newman
for a distant observer  \cite{burhild,me2006}.
 Rotation transforms the de Sitter center into de Sitter vacuum disk
  which has properties of a perfect conductor and ideal
diamagnetic and displays superconducting behavior within a single
spinning object \cite{me2006,portrait}.

In this paper we address the question of the existence and
asymptotic behaviour of solutions to the dynamical field equations
which define the basic generic features of regular rotating
electrically charged objects. In the paper \cite{behm} it was noted that
rotation leads to violation of the weak energy condition (WEC) in
interior regions of neutral regular rotating configurations.
Violation of WEC was found for regular solutions obtained with the
Newman-Janis algorithm from the Hayward and Bardeen metrics
\cite{bambi}, for rotating regular solutions obtained by
postulating a metric $g_{\mu\nu}$ and calculating $T_{\mu\nu}$
from the Einstein equations \cite{neves}, and for the
Ay\'on-Beato-Garcia solution transformed to rotational form with
the Newman-Janis algorithm \cite{stuchlik}. Here we study the weak
energy condition for NED-GR electrically charged rotating regular
structures with an arbitrary Lagrangian ${\cal L}(F)$, and generic
behaviour of electromagnetic fields on the vacuum surfaces
($p+\rho=0$), beyond which the weak energy condition could be
violated.

The paper is organized as follows. In Sect. 2 we present and
analyze the basic equations. In Sect. 3 we address the question of
the existence of horizons and ergoregions, and present interior de
Sitter vacuum surfaces. In Sect. 4 we study  electromagnetic
fields and in Sect 5 behaviour of fields in regular interiors. In
Sect. 6 we summarize and discuss the results.

\section{Basic equations}

Nonlinear electrodynamics minimally coupled to gravity is
described by the action
 \beq
 S=\frac{1}{16\pi G}\int{d^4
x\sqrt{-g}[R-{\cal L}(F)]}; ~~ ~ F=F_{\mu\nu}F^{\mu\nu} ,
                                                                          \label{action}
  \eeq
where $R$ is the scalar curvature, and
$F_{\mu\nu}=\partial_{\mu}A_{\nu}-\partial_{\nu}A_{\mu}$ is the
electromagnetic field. The gauge-invariant electromagnetic
Lagrangian ${\cal L}(F)$ is an arbitrary function of $F$ which
should have the Maxwell limit, ${\cal L} \rightarrow F,~ {\cal
L}_F\rightarrow 1$ in the weak field regime.

Variation with respect to $A^{\mu}$ and $g_{\mu\nu}$ yields
 the dynamic field equations
  \beq
\nabla_{\mu}({\cal L}_F F^{\mu\nu})=0;
                                                                          \label{DynEq}
  \eeq
  \beq
 \nabla_{\mu}{^*}F^{\mu\nu}=0;~~
^{\star}F^{\mu\nu}=\frac{1}{2}\eta^{\mu\nu\alpha\beta}F_{\alpha\beta};
~~ \eta^{0123}=-\frac{1}{\sqrt{-g}} ,
                                                                                \label{BianEq}
 \eeq
  where greek indices run from 0 to 3 and ${\cal L}_F=d{\cal L}/dF$,
and the Einstein equations $G_{\mu\nu}=-\kappa T_{\mu\nu}$ with
the stress-energy tensor
  \beq
   \kappa T^{\mu}_{\nu}=-2{\cal L}_F
F_{\nu\alpha}F^{\mu\alpha}+\frac{1}{2}\delta^{\mu}_{\nu}{\cal
L};~~~\kappa=8\pi G .
                                                                                   \label{set}
  \eeq
NED-GR equations do not admit regular spherically symmetric
solutions with the Maxwell center \cite{kirill}, but they  admit
regular solutions with the de Sitter center  in which field
tension goes to zero, while the energy density of the
electromagnetic vacuum $T_t^t$ achieves its maximal finite value
which represents the de Sitter cutoff on the self-energy density
\cite{me2004}. The question of correct description of NED-GR
regular electrically charged structures by the Lagrange dynamics
is clarified in \cite{us2015}. Regular spherical solutions
satisfying (\ref{myvac}) are described by the metric
 \beq
 ds^2=g(r) dt^2 - \frac{dr^2}{g(r)} - r^2 d\Omega^2;
 ~~g(r)=1-\frac{2{\cal M}(r)}{r}; ~~
 {\cal
M}(r)=4\pi\int_0^r{~\tilde\rho(x)x^2dx} ,
                                                                                     \label{spherMetric}
  \eeq
 with the electromagnetic density $\tilde\rho(r)=T^t_t(r)$ from
(\ref{set}). This metric has the de Sitter asymptotic as
$r\rightarrow 0$ and the Reissner-Nordstr\"om asymptotic as
$r\rightarrow\infty$ \cite{me2004}.

Spherically symmetric solutions of the Einstein equations
specified by (\ref{myvac}) belong to the Kerr-Schild class
\cite{ssqv,behm}. By the G\"urses-G\"ursey algorithm they can be
transformed into axially symmetric regular solutions describing
rotating objects \cite{gurses}.  The Kerr-Schild metric has the
form
   \beq
   g_{\mu\nu} = \eta_{\mu\nu}
+ \frac{2f(r)}{\Sigma}k_{\mu} k_{\nu} ,
                                                                                 \label{ksmetric}
\eeq
 where $\eta_{\mu\nu}$ is the Minkowski metric and $k_{\mu}$
is a vector  field tangent to the Kerr principal null
congruence. Metric (\ref{ksmetric}) involves a function
$f(r)=r{\cal M}(r)$ which comes from a spherically symmetric
solution \cite{gurses}. For the Kerr-Newman geometry $f(r) = mr -
e^2/2$. The
parameter $r$ is defined as an affine parameter along either of
two principal null congruences. The surfaces of constant $r$ are
the oblate confocal ellipsoids of revolution given by
\cite{chandra}
 \beq
r^4-(x^2+y^2+z^2-a^2)r^2-a^2 z^2=0
                                                                                \label{ellipsoid}
\eeq
 which degenerate, for $r = 0$, to the equatorial disk
\beq
 x^2 + y^2 \leq a^2, ~~ z = 0
                                                                              \label{disk}
\eeq
 centered on the symmetry axis and bounded by the ring
\beq
 x^2 + y^2 = a^2, ~ ~ z = 0 .
                                                                                 \label{ring}
\eeq
In the Kerr-Newman metric (\ref{kn}) the ring (\ref{ring}) comprises the Kerr singularity of the Kerr-Newman
geometry \cite{chandra}.

The Cartesian coordinates $x, y, z$ are related to the
Boyer-Lindquist coordinates $r, \theta, \phi$ by
 \beq
 x^2 + y^2 =
(r^2 + a^2)\sin^2\theta;~~z=r\cos\theta .
                                                                              \label{coordinates}
\eeq
 In the Boyer-Lindquist coordinates the G\"urses-G\"ursey metric
 reads
  \beq
  ds^2 = \frac{2f - \Sigma}{\Sigma} dt^2  +
\frac{\Sigma}{\Delta} dr^2 + \Sigma d\theta^2 -
\frac{4af\sin^2\theta}{\Sigma}dt d\phi + \biggl(r^2 + a^2 +
\frac{2fa^2\sin^2\theta}{\Sigma}\biggr)\sin^2\theta d\phi^2 ,
                                                                                  \label{metric}
\eeq
 where the Lorentz signature is [- + + +], and
\beq
 \Delta(r) = r^2 + a^2 - 2f(r); ~  ~\Sigma=r^2+a^2\cos^2\theta .
                                                                                \label{delta}
  \eeq
For the Kerr-Newman geometry $2f(r) = 2mr - e^2$ can change the sign which leads
to causality violation related to regions where $g_{\phi\phi} < 0$.
For regular spherical solutions satisfying the
weak energy condition, $f(r)=r{\cal M}(r)$ is non-negative
function since ${\cal M}(r)$ is monotonically growing from
$4\pi\tilde\rho(0)r^3/3$ as $r\rightarrow 0$ to $m-e^2/2r$ as
$r\rightarrow\infty$ \cite{me2004}. This guarantees the causal
safety on the whole manifold due to $f(r)\geq 0$ and $g_{\phi\phi}
> 0$ in (\ref{metric}).  For regular spherical configurations
$\kappa(p_{\perp}+\rho)=-{\cal L}_FF$,  the field invariant $F$ is
non-positive function evolving from $F=-0$ as $r\rightarrow 0$ to
$F=-0$ for $r\rightarrow\infty$ \cite{me2004}, ${\cal L}_F$ plays
the role of the electric permeability \cite{burhild,me2004}, and
electrodynamics of the continued media requires ${\cal L}_F > 0$
\cite{landau2}, so that WEC is always satisfied. The mass
parameter $m=4\pi\int_0^{\infty}{{\tilde\rho}(r)r^2dr}$ appearing
in a spinning solution, is the finite positive electromagnetic
mass \cite{me2004}, generically related to interior de Sitter
vacuum for any solution from the class specified by (\ref{myvac})
\cite{me2002}.

 The anisotropic stress-energy tensor responsible for
(\ref{metric})  can be written in the form \cite{gurses}
     \beq
 T_{\mu\nu}=(\rho + p_{\perp})(u_{\mu}u_{\nu} -
l_{\mu}l_{\nu}) + p_{\perp} g_{\mu\nu}
                                                                              \label{GGset}
  \eeq
in the orthonormal tetrad
 \beq
u^{\mu} = \frac{1}{\sqrt{\pm\Delta\Sigma}}[(r^2 + a^2)
\delta^{\mu}_0 + a \delta^{\mu}_3], ~~ l^{\mu} =
\sqrt{\frac{\pm\Delta}{\Sigma}}\delta^{\mu}_1,  ~~ n^{\mu} =
\frac{1}{\sqrt{\Sigma}}\delta^{\mu}_2,  ~~ m^{\mu} =
\frac{-1}{\sqrt{\Sigma}\sin\theta}[a\sin^2\theta \delta^{\mu}_0 +
\delta^{\mu}_3] .
                                                                              \label{tetrad}
 \eeq
 The sign plus refers to the regions outside the event horizon
 and inside the Cauchy horizon where the vector $u^{\mu}$ is time-like,
 and the sign minus refers to
 the regions between the horizons where the vector $l^{\mu}$ is
 time-like. The vectors $m^{\mu}$ and $n^{\mu}$ are space-like in
 all regions.

 The eigenvalues of the stress-energy tensor (\ref{set}) in the co-rotating frame
 where each of ellipsoidal layers rotates with the angular velocity
$\omega(r) = u^{\phi}/u^t = a/(r^2 + a^2)$ \cite{behm}, are
defined by
  \beq
  T_{\mu\nu}u^{\mu}u^{\nu} = \rho(r, \theta);
~~T_{\mu\nu}l^{\mu}l^{\nu} = p_r = - \rho;  ~~
T_{\mu\nu}n^{\mu}n^{\nu} = T_{\mu\nu}m^{\mu}m^{\nu} = p_{\perp}(r,
\theta) .
                                                                           \label{set-eigenvalues}
 \eeq
in the regions outside the event horizon and inside the Cauchy
horizon where density is defined as the eigenvalue of the
time-like eigenvector $u^{\mu}$. They are related to the function
$f(r)$ as \cite{behm}
  \beq
 \kappa\Sigma^2\rho = {2(f'r - f); ~ ~\kappa\Sigma^2 p_{\perp}
= 2(f'r - f) - f^{\prime\prime}\Sigma}  .
                                                                       \label{rho-pressure}
 \eeq
It follows
  \beq
\kappa\rho=\frac{r^4}{\Sigma^2}{\tilde\rho}(r);~~ \kappa
p_{\perp}=\left(\frac{r^4}{\Sigma^2}-\frac{2r^2}{\Sigma}\right)
\tilde{\rho}(r) -\frac{r^3}{2\Sigma}{\tilde\rho}^{\prime}(r) .
                                                                      \label{setcomp}
  \eeq
The prime denotes the derivative with respect to $r$. In the
co-rotating frame we thus have
 \beq
\kappa(p_{\perp}+\rho)=2\left(\frac{r^4}{\Sigma^2}-\frac{r^2}{\Sigma}\right)
{\tilde\rho}(r)-\frac{r^3}{2\Sigma}{\tilde\rho}^{\prime}(r) .
                                                                        \label{press-dens}
  \eeq

The basic features of regular rotating objects  follow from
generic behaviour of related regular spherical solutions without
specifying the particular form of the NED lagrangian ${\cal
L}(F)$.

\section{Geometry}

\subsection{Horizons and ergospheres}

Horizons are defined by zeros of the function $\Delta(r)$ in (\ref{delta}) which can be written as
 \beq
\Delta(r)=r^2 + a^2 - 2f(r) = a^2+r^2g(r) .
                                                                             \label{deltag}
  \eeq
$\Delta=a^2$ at zero points of the metric function $g(r_{h})=0$,
and evolves from $\Delta=a^2$ as $r=0$ to
$\Delta\rightarrow\infty$ as $r\rightarrow\infty$.

The number of horizons depends on the generic properties of the
metric function $g(r)$, which has at most two zero points and one
minimum  \cite{me2002}.

{\bf H1. The case of two zero points of the metric function
$g(r)$.} Derivatives of $\Delta$ are
  \beq
\Delta^{\prime}=2rg(r)+r^2g^{\prime}(r); ~~
\Delta^{\prime\prime}
=2g(r)+4rg^{\prime}(r)+r^2g^{\prime\prime}(r) .
                                                                         \label{deltaderiv}
  \eeq
At $r=0$ derivatives take the values
$\Delta^{\prime}=0;~\Delta^{\prime\prime}=2$ and the function
$\Delta$ has the minimum, $\Delta=a^2$. Next it grows and can have
maximum at a certain value $r_m < r_{a}$ where $r_{a}$ is the
first zero of $g(r)$. At the maximum  $\Delta^{\prime}(r_m)=0$ and
hence $g^{\prime}(r_m)=-2g(r_m)/r_m$, second derivative
$\Delta^{\prime\prime}(r_m)=-6g(r_m)+r_m^2g^{\prime\prime}(r_m)<0$.
After passing the maximum $\Delta$ achieves $\Delta=a^2$ at the
first zero of $g(r)$, then it will achieve this value at the
second zero point $r_{b}$ of $g(r)$. Between $r_{a}$ and $r_{b}$
it has at least one extremum, which is the minimum, because in the
region $r_{a}<r<r_{b}$ the metric function $g(r)$ is negative and
has the minimum. In this region $g^{\prime}$ is first negative,
then passes zero and becomes positive, hence $g^{\prime\prime}\geq
0$ everywhere between zeros of $g(r)$, while $g\leq 0$, as a
result $\Delta^{\prime\prime}\geq 0$ everywhere in the considered
region. It is evident that in this region the function $\Delta(r)$
can have only minimum and only one.

At $r>r_{b}$ we have $g(r)>0$ and $g^{\prime}>0$ so that $\Delta$
cannot vanish. We proved that in the case of two zero points of
the metric function $g(r)$ the number of zero points of the
function $\Delta(r)$ is maximum two, i.e. axially symmetric
spacetime can have at most two horizons.

\vskip0.1in

{\bf H2. The case of the double root of the function $g(r)$.} In this
case $g(r)\geq 0$ everywhere. $\Delta^{\prime}=0$ at
$r_{a}=r_{b}=r_d$. In this point $g^{\prime\prime}>0$ and
$\Delta^{\prime\prime}>0$, hence $\Delta$ has the minimum
$\Delta=a^2$. In this case $\Delta(r)$ is everywhere positive function which has one maximum at $r=0$
and one minimum at $r=r_d$. Axially
symmetric spacetime does not have horizons.

{\bf H3. The  case when $g(r)>0$ everywhere}  does not differ
essentially from the case {\bf B}.  The function $\Delta(r)$ is
positive everywhere. Extremum of $\Delta$ can be only in the
region where $g^{\prime} < 0$, but in this case it can  have an inflection
point instead of an extremum.

{\bf Ergospheres and ergoregions.} Ergosphere is a surface of a
static limit $g_{tt}=0$ given by
 \beq
 g_{tt}(r, \theta)=r^2 + a^2\cos^2\theta - 2f(r)=0
                                                                             \label{ergosphere}
  \eeq
It follows that $z^2=(2r^2f(r)-r^4)/a^2$. Each point of the
ergosphere belongs to some of confocal ellipsoids
(\ref{ellipsoid}) covering the whole space as the coordinate
surfaces $r$=const. The width of the ergosphere at a certain $z$
is $x^2+y^2=(a^2+r^2)(1-z^2/r^2)$. In the equatorial plane
$x^2+y^2=a^2+r^2$ provided that $z^2=0=2f(r)-r^2$. For any regular
density profile the function $f(r)$ is everywhere positive and
monotonically grows from $Dr^4$ as $r\rightarrow0$, where $D$ is
constant, to $mr$ where $m$ is the mass parameter. Ergosphere
exists when the curve $u=2f(r)$ intersects or touches the parabola
$u=r^2$ (curve 2 in Fig.1). It is evident that in this case the curve $u=2f(r)$
intersects also the (situated above) parabola
$u=r^2+a^2\cos^2\theta$ for a given $\theta$. There are four cases
of the existence of ergospheres and ergoregions (the regions where
$g_{tt} < 0$).

\vskip0.1in

{\bf E1. Black hole case.} In this case, ergospheres and
ergoregions exist for any density profile. At  $z$-axis
equations (\ref{deltag}) and (\ref{ergosphere}) are identical, so
that the minor axis of the ergosphere is equal $r_+$. In accordance
with (\ref{deltag}), the function $u=2f(r)$ intersects or touches
the parabola $u=r^2+a^2$ (curve 1 in Fig.1) and hence intersects the (situated below)
parabola $u=r^2$, since near $r\rightarrow 0$
a function $f(r)$ goes to zero as $r^4$, faster than $r^2$. In the
case of two horizons  the curve $u=2f(r)$ intersects the parabola
$u=r^2+a^2$ on the internal horizon $r=r_-$, then on the event
horizon $r=r_+$ and ultimately at the point $r=r_e$ which defines
the width of the ergosphere in the equatorial plane
$x^2+y^2=a^2+r_e^2$. In this case the ergoregion  exists between
the event horizon $r=r_+$ and the ergosphere (curve 3a in
Fig.1). In the case of a double horizon, the ergoregion exists
between $r=r_{\pm}$ and the ergosphere (curve 3b in Fig.1).

%FIGURE1
\begin{figure}[htp]
%%\vspace{-8.0mm}
\centering
\epsfig{file=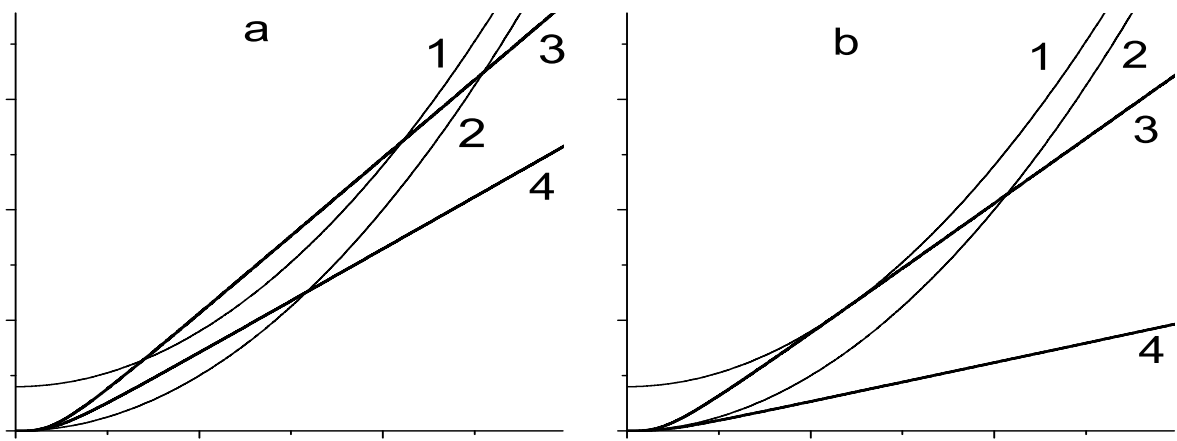,width=17.3cm,height=7.1cm}
\caption{Functions $2f(r)$ for the four cases of existence of
ergosphere.} \label{fig.1}
\end{figure}

{\bf E2. Soliton case.} In the case of absence of horizons, there
are three options. The curve $u=2f(r)$ can intersect the parabola
$u=r^2$ twice, and there exists ergoregion between two surfaces
of intersection (curve 4a). Second option is that the curve
$u=2f(r)$ touches the parabola $u=r^2$ in a certain point $r=r_e$ (curve 4b).
In this case ergoregion exists beyond the ergosphere whose width
in the equatorial plane is $x^2+y^2=a^2+r_e^2$ and includes the
whole interior. Third option is no intersections or touching, and
hence no ergospheres.

\subsection{De Sitter vacuum surfaces}

Near the disk (\ref{disk}) the function $f(r)$ in (\ref{metric})
approaches de Sitter asymptotic \cite{me2004}
  \beq
  2f(r)\rightarrow
\frac{\kappa{\tilde\rho(0)}}{3}r^4=\frac{r^4}{r_0^2};
~~~r_0^2=\frac{3}{\kappa{\tilde
\rho(0)}} .
                                                                     \label{f-on-disk}
    \eeq
With taking into account $\Sigma=(r^4+a^2 z^2)r^{-2}$, we get
 \beq
\frac{2f(r)}{\Sigma}=\frac{r^4}{r_0^2}\frac{r^2}{(r^4+a^2
z^2)} .
                                                                          \label{f-Sigma}
 \eeq
In the equatorial plane it gives
${2f(r)}/{\Sigma}=\kappa\tilde\rho(0){r^2}/3$, so that the disk
$r=0$ is intrinsically flat. But vacuum density is non-zero
throughout the whole disk. In the equatorial plane
${r^2}/{\Sigma}\rightarrow 1$ as $z\rightarrow 0$
\cite{me2006}, and the equation (\ref{press-dens}) reduces to
  \beq
\kappa(p_{\perp}+\rho)=-{r}{\tilde\rho}^{\prime}(r)/2 .
                                                                          \label{ursosekwator}
  \eeq
  For the regular spherical solutions $\tilde\rho^{\prime}\leq 0$
\cite{me2002}, and the weak energy condition is satisfied for
axially symmetric solutions in the equatorial plane.

By (\ref{setcomp}), the density in the equatorial plane is
$\rho(r,\theta)={\tilde\rho}(r)$. When $r\rightarrow 0$,
${\tilde\rho}(r)\rightarrow \tilde\rho(0)$, so that on the disk
$\rho(r,\theta)={\tilde\rho}(0)$. For the spherically symmetric
solutions regularity requires
$r{\tilde\rho}^{\prime}(r)\rightarrow 0$ as $r\rightarrow 0$
\cite{me2004}. As a result the equation (\ref{ursosekwator}) gives
on the disk the  equation of state
     \beq
 p_{\perp}=p_r=-\rho
                                                                      \label{press-on-disk}
\eeq
 which represents the rotating  de Sitter vacuum.

The equation (\ref{press-dens}) can be written as
   \beq
   \kappa (p_{\perp}+\rho)=\frac{2r^2}{\Sigma^2}\left(\frac{\Sigma
r}{4}|\tilde\rho^{\prime}|
  -{\tilde\rho}a^2\cos^2\theta\right).
                                                                                      \label{VIP}
  \eeq
It implies a possibility of generic violation of the weak energy
condition. WEC requires $T_{\mu\nu}w^{\mu}w^{\nu}\geq 0$ for any
time-like vector $w^{\mu}$. Representing vector $w^{\mu}$  in the
tetrad (\ref{tetrad}) we find that WEC is valid if $\rho\geq 0$
and $p_{\perp}+\rho\geq 0$. The first condition is satisfied
according to (\ref{setcomp}). WEC can be thus violated beyond the
vacuum surface ${\cal E}(r, z)=0$ on which $p_{\perp}+\rho=0$ and
the right-hand side in (\ref{VIP}) can change its sign. It can be
expressed through the pressure of a related spherical solution,
$~\tilde{p_{\perp}}=-\tilde\rho-r\tilde\rho^{\prime}/2~$
\cite{me2002}, which gives
  \beq
  \kappa(p_{\perp}+\rho)=\frac{r|\tilde\rho^{\prime}|}{2\Sigma^2}~{\cal
  E}(r,z)=0;~~{\cal
  E}(r,z)=r^4-z^2P(r);~~P(r)
  =\frac{2a^2}{r|\tilde\rho^{\prime}|}(\tilde\rho
  -\tilde{p_{\perp}}) .
                                                                         \label{e-surface}
 \eeq
As we see, the existence of vacuum surfaces and hence possible
violation of the weak energy condition is possible only for the
mass functions originated from spherical solutions satisfying  the
dominant energy condition ($\tilde\rho \geq \tilde{p_k}$).

%%%%%%%%%%%%%%%%%%%%%%%%%%%%%%%%%%%%%%%%%%%%%%%%%%%%%
Each point of the ${\cal E}$-surface belongs to some of confocal
ellipsoids (\ref{ellipsoid}). In the Cartesian coordinates
(\ref{coordinates}) the equation of the ellipsoid
(\ref{ellipsoid}) reads $(r^2-z^2)(a^2+r^2)=r^2(x^2+y^2)$. On the
${\cal E}$-surface we have $z^2=r^4/P(r)$. The squared width of
the ${\cal E}$-surface, $W^2_{\cal E}=(x^2+y^2)_{\cal E} =
(a^2+|z|\sqrt{P(r)})(1-|z|/\sqrt{P(r)})$. It is easily to show
that ${\cal E}$-surface is entirely confined within the
$r_*$-ellipsoid whose minor axis coincides with $|z|_{max}$ for
the ${\cal E}$-surface \cite{interior}.  For regular solutions
$r\tilde\rho^{\prime}\rightarrow 0$,~ $p_{\perp}\rightarrow -\rho$
as $r\rightarrow 0$ \cite{me2004}, and $P(r)\rightarrow
A^2r^{-(n+1)}$ as $r\rightarrow 0$, with the integer $n\geq 0$ .
As a result the derivative of $W_{\cal E}(z)$ near $z\rightarrow
0$ behaves as $z^{-(n+1)/(n+5)}$ and goes to $\pm\infty$ as
$z\rightarrow 0$, so that the function $W_{\cal E}(z)$ has the
cusp at approaching the disk and at least two symmetric (with
respect to the equatorial plane) maxima  between $z=\pm r_*$ and
$z=0$ \cite{interior}.

The $\cal E$-surface is plotted below for the
regularized Coulomb profile \cite{me2004}
  \beq
  \tilde\rho=\frac{q^2}{(r^2+r_q^2)^2}; ~~
  r_q=\frac{\pi}{8}\frac{q^2}{m} .
                                                                            \label{profile}
 \eeq
For this profile $P(r_*)=r_*^2=r_qa$, and the ${\cal E}$-surface
is given by ${\cal E}(r, z)= r^6-r_*^4z^2=0$.

The width of the ${\cal E}$-surface $W_{\cal E}=(x^2+y^2)_{\cal
E}$ as the function of $z$ has two maxima at $z_m=\pm
r_m^3/r_*^2$. Relation between the width of the ${\cal E}$-surface
in the equatorial plane $W_{\cal E}=a$  and its height
$|z|_{max}=\sqrt{ar_q}$  defines the explicit form of the ${\cal
E}$-surface. It depends on two parameters: $\alpha=a/m$ and
specific charge $\beta=q/m$. In terms of these parameters
  \beq
H_{{\cal
E}}=\sqrt{\frac{a}{r_q}}=\frac{2\sqrt{2\alpha}}{\sqrt{\pi}\beta}; ~~
~W_{\cal E}=\frac{a}{r_q}=\frac{8\alpha}{\pi\beta^2} .
                                                                                \label{sizes}
 \eeq
For $\alpha < \pi\beta^2/8$ we have $H_{\cal E}/W_{\cal E} > 1$
and the ${\cal E}$-surface is prolate (Fig. 2). For  black holes
the parameter $\beta$ changes within the range $0 < \beta < 0.99$
\cite{cardoso,hamilton}. It can be the case of a slowly rotating
moderately charged black hole.
%FIGURE2
\begin{figure}[htp]
%%\vspace{-8.0mm}
\centering
\epsfig{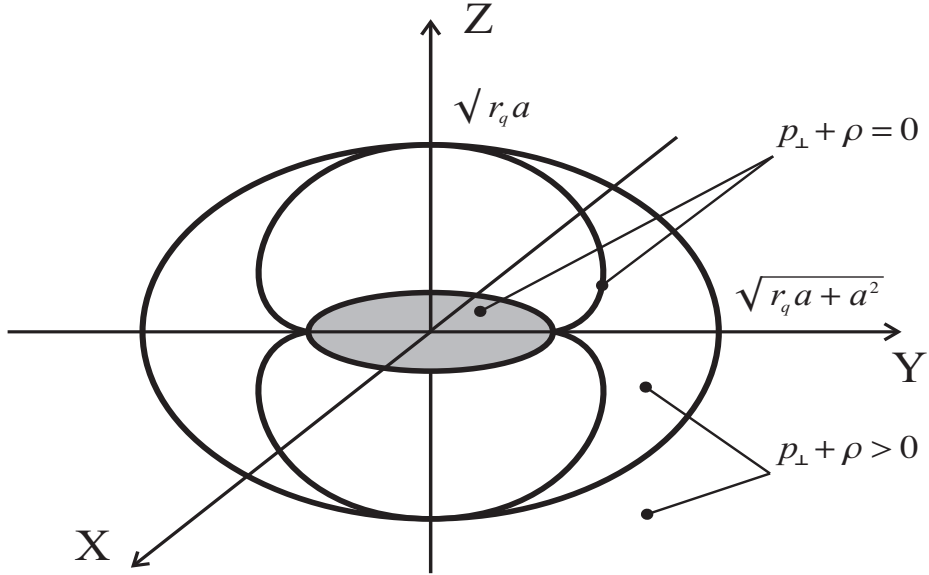}
\caption{Vacuum ${\cal E}$-surface for the case $\alpha <
\pi\beta^2/8$.} \label{fig.2}
\end{figure}

In the case $\alpha > \pi\beta^2/8$, the ${\cal E}$-surface is
oblate (Fig. 3). It is  the case for electromagnetic spinning
soliton. It can be also the case of a slightly charged rotating
black hole and of an extreme black hole.
%FIGURE3
\begin{figure}[htp]
%%\vspace{-8.0mm}
\centering \epsfig{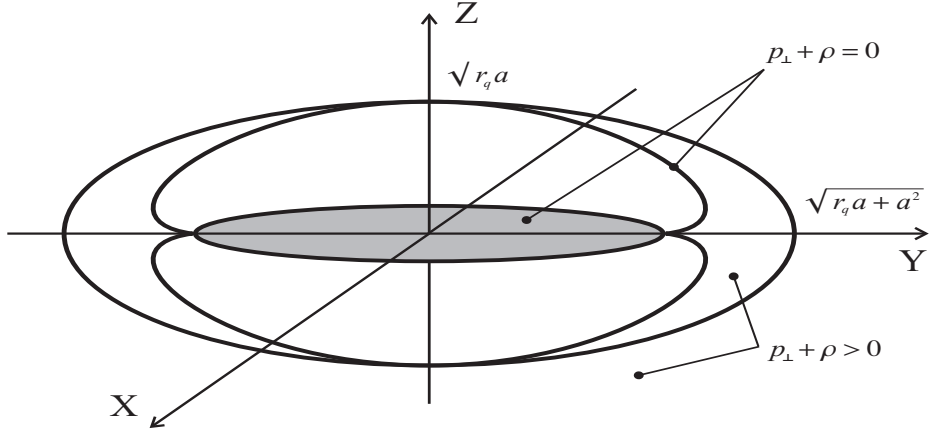}
\caption{Vacuum ${\cal E}$-surface for the  case $\alpha
> \pi\beta^2/8$.} \label{fig.3}
\end{figure}

\section{Electromagnetic fields}

\subsection{Field equations}

Non-zero field components compatible with the axial symmetry are
$F_{01}, F_{02}, F_{13}, F_{23}$. In geometry with the metric
(\ref{metric}) they are related by
  \beq
F_{31}=a\sin^2\theta F_{10}; ~~ aF_{23}=(r^2+a^2)F_{02} .
                                                                        \label{FieldComp}
  \eeq
The field invariant $F=F_{\mu\nu}F^{\mu\nu}$ in the axially
symmetric case reduces to
  \beq
F=2\left(\frac{F_{20}^2}{a^2\sin^2\theta}-F_{10}^2\right) .
                                                                      \label{AxInvariant}
 \eeq

%%%%%%%%%%%%%%%%%%%%%%%%%%%%%%%%%%%%%%%%

In terms of the  3-vectors, denoted by the italic indices running
from 1 to 3 and defined as
   \beq
{E}_j=\{F_{j0}\}; ~~ {D}^j=\{{\cal L}_F F^{0j}\};~
~{B}^j=\{{^*}F^{j0}\};
 ~~ {H}_j=\{{\cal L}_F {^*}F_{0j}\} ,
                                                                            \label{fields}
  \eeq
the field equations (\ref{DynEq})-(\ref{BianEq}) take the form of
the source-free Maxwell equations
  \beq
\nabla{\mathbf{D}}=0;~~
\nabla\times{\mathbf{H}}=\frac{\partial{\mathbf{D}}}{\partial t};~
~ ~  \nabla{\mathbf{B}}=0;~~
\nabla\times{\mathbf{E}}=-\frac{\partial{\mathbf{B}}}{\partial t} .
                                                                      \label{maxwell}
  \eeq

The electric induction  ${\mathbf{D}}$ and the magnetic induction
$\mathbf{B}$ are related with
 the electric and magnetic field intensities by
   \beq
D^{j}=\epsilon^{j}_{k}E^{k}; ~~ B^{j}=\mu^{j}_{k}H^{k} ,
                                                                                 \label{s2}
 \eeq
where $\epsilon_{j}^{k}$ and $\mu_{j}^{k}$ are the tensors of the
electric and magnetic permeability given by \cite{me2006}
  \beq
\epsilon_r^r=\frac{(r^2+a^2)}{\Delta}{\cal L}_F;~~
\epsilon_{\theta}^{\theta}={\cal L}_F; ~~
\mu_r^r=\frac{(r^2+a^2)}{\Delta{\cal L}_F}; ~~
\mu_{\theta}^{\theta}=\frac{1}{{\cal L}_F}  .
                                                                                   \label{s4}
  \eeq

%%%%%%%%%%%%%%%%%%%%%%%%%%%%%%%%%%%%%%%%%%%%%%%%%%%%%%%%%%%%%%

\newpage

Equations (\ref{DynEq}) and (\ref{BianEq}) form the system
   \beq
   \frac{\partial}{\partial r}[(r^2+a^2)\sin\theta {\cal L}_F
   F_{10}]+\frac{\partial}{\partial\theta}[\sin\theta {\cal L}_F
   F_{20}]=0
                                                                            \label{f1}
  \eeq
  \beq
  \frac{\partial}{\partial r}\left[\frac{1}{\sin\theta}{\cal L}_F
  F_{31}\right]+\frac{\partial}{\partial\theta}
  \left[\frac{1}{(r^2+a^2)\sin\theta}{\cal
  L}_F F_{32}\right]=0
                                                                              \label{f2}
 \eeq
   \beq
\frac{\partial F_{23}}{\partial r}+\frac{\partial
F_{31}}{\partial\theta}=0
                                                                              \label{f3}
  \eeq
  \beq
  \frac{\partial F_{01}}{\partial\theta}+\frac{\partial
  F_{20}}{\partial r}=0 .
                                                                             \label{f4}
  \eeq
Four field components are related by two equations
(\ref{FieldComp}), so only two are independent and we can
apply (\ref{FieldComp}) to transform this
 system to the form
 \beq
 \frac{\partial}{\partial r}\biggl[(r^2+a^2)\sin\theta{\cal
 L}_FF_{10}\biggr]
 +\frac{\partial}{\partial\theta}\biggl[\sin\theta{\cal L}_FF_{20}\biggr]=0
                                                                                 \label{sep1}
 \eeq
 \beq
 \frac{\partial}{\partial r}\biggl[a\sin\theta {\cal
 L}_FF_{10}\biggr]
 +\frac{\partial}{\partial\theta}\biggl[\frac{1}{a\sin\theta}{\cal
 L}_FF_{20}\biggr]=0
                                                                              \label{sep2}
\eeq
  \beq
 \frac{\partial}{\partial
 r}F_{20}-\frac{\partial}{\partial\theta}F_{10}=0
                                                                              \label{sep3}
    \eeq
    \beq
\frac{\partial}{\partial\theta}\biggl[a^2\sin^2\theta
F_{10}\biggr]-\frac{\partial}{\partial
r}\biggl[(r^2+a^2)F_{20}\biggr]=0 .
                                                                              \label{sep4}
   \eeq
   For calculation of derivatives in (\ref{sep1})-(\ref{sep2}) we need derivatives of the
   invariant $F$ which read
 \beq
 \frac{\partial F}{\partial
 r}=4\biggl(\frac{F_{20}}{a^2\sin^2\theta}\frac{\partial F_{20}}{\partial
 r}-F_{10}\frac{\partial F_{10}}{\partial
 r}\biggr)
                                                                               \label{Fr}
    \eeq
    \beq
    \frac{\partial F}{\partial\theta}
    =4\biggl(\frac{F_{20}}{a^2\sin^2\theta}\frac{\partial
    F_{20}}{\partial\theta}-\frac{\cot\theta}{a^2\sin^2\theta}F_{20}^2
    -F_{10}\frac{\partial F_{10}}{\partial\theta}\biggr) .
                                                                               \label{Ft}
     \eeq
With taking into account (\ref{Fr}), (\ref{Ft}) and (\ref{sep3}),
we reduce the system (\ref{sep1})-(\ref{sep4}) to
  $$
  (r^2+a^2)\biggl\{{\cal L}_F-4{\cal L}_{FF}F_{10}^2\biggr\}\frac{\partial
F_{10}}{\partial r} + \biggl\{{\cal L}_F+ 4{\cal
L}_{FF}\frac{F_{20}^2}{a^2\sin^2\theta}\biggr\}\frac{\partial
F_{20}}{\partial\theta}
    $$
     \beq
     +4{\cal
L}_{FF}\Sigma\frac{F_{10}F_{20}}{a^2\sin^2\theta}\frac{\partial
F_{20}}{\partial r} + 2r{\cal L}_FF_{10}
+F_{20}\cot\theta\biggl\{{\cal L}_F-4{\cal
L}_{FF}\frac{F_{20}^2}{a^2\sin^2\theta}\biggr\}=0
                                                                               \label{sep5}
   \eeq
   \beq
\biggl\{{\cal L}_F-4{\cal L}_{FF}F_{10}^2\biggr\}\frac{\partial
F_{10}}{\partial r}+\frac{1}{a^2\sin^2\theta}\biggl\{{\cal
L}_F+4{\cal
L}_{FF}\frac{F_{20}^2}{a^2\sin^2\theta}\biggr\}\frac{\partial
F_{20}}{\partial\theta}-\frac{\cot\theta}{a^2\sin^2\theta}F_{20}
\biggl\{{\cal L}_F + 4{\cal
L}_{FF}\frac{F_{20}^2}{a^2\sin^2\theta}\biggr\}=0
                                                                              \label{sep6}
  \eeq
  \beq
  \Sigma\frac{\partial F_{20}}{\partial
  r}+2rF_{20}-a^2\sin{2\theta}F_{10}=0
                                                                              \label{sep7}
    \eeq
    \beq
    \frac{\partial}{\partial
    r}F_{20}-\frac{\partial}{\partial\theta}F_{10}=0 .
                                                                            \label{sep8}
\eeq
    Two of these equations are strongly non-linear.

\subsection{Asymptotic solutions}

Dynamical equations $\nabla_{\mu}({\cal L}_FF^{\mu\nu})=0$ are
satisfied by the functions \cite{me2006}
  \beq
  \Sigma^2({\cal L}_FF_{01})=-q(r^2-a^2\cos^2\theta);~
  ~\Sigma^2({\cal L}_FF_{02})=qa^2r\sin 2\theta;
                                                                                 \label{f6}
   \eeq
    \beq
  \Sigma^2({\cal L}_FF_{31})=aq\sin^2\theta(r^2-a^2\cos^2\theta); ~~
  ~\Sigma^2({\cal L}_FF_{23})=aqr(r^2+a^2)\sin 2\theta .
                                                                                    \label{f5}
  \eeq
In the case  ${\cal L}_F=1$, they satisfies also the dynamical
equations (\ref{BianEq}) and coincide with the solutions to the
Maxwell-Einstein equations \cite{carter,tiomno}.

The field functions (\ref{f5})-(\ref{f6}) satisfy the equations
  \beq
  \frac{\partial ({\cal L}_F F_{23})}{\partial
r}+\frac{\partial ({\cal L}_F F_{31})}{\partial\theta}=0
                                                                                   \label{f3A}
  \eeq
  \beq
  \frac{\partial({\cal L}_F F_{01})}{\partial\theta}+\frac{\partial
  ({\cal L}_FF_{20})}{\partial r}=0 .
                                                                                          \label{f4A}
  \eeq
It follows
  \beq
\frac{\partial F_{23}}{\partial r}+\frac{\partial
F_{31}}{\partial\theta}=-\frac{{\cal L}_{FF}}{{\cal
L}_F}\left[F_{31}\frac{\partial F}{\partial
\theta}+F_{23}\frac{\partial F}{\partial r}\right]
                                                                                          \label{f3B}
  \eeq
  \beq
  \frac{\partial F_{01}}{\partial\theta}+\frac{\partial
  F_{20}}{\partial r}=-\frac{{\cal L}_{FF}}{{\cal
L}_F}\left[F_{01}\frac{\partial F}{\partial
\theta}+F_{20}\frac{\partial F}{\partial r}\right]
                                                                                       \label{f4B}
  \eeq
Left sides vanish identically when right sides are zero. This
defines the following cases when the functions
(\ref{f5})-(\ref{f6}) satisfy the dynamical system
(\ref{f1})-(\ref{f4}):

{\bf (A)} ${\cal L}_{FF}=0, {\cal L}_{F}\neq 0$, the case of the
linear electromagnetic field;

{\bf (B)} ${\cal L}_F=\infty, {\cal L}_{FF}\neq 0$, strongly
nonlinear regime. Density of electromagnetic energy given by
(\ref{setcomp}) grows towards the interior disk $r=0$. Applying
(\ref{f5})-(\ref{f6}) we obtain
  \beq
\kappa{(p_{\perp}+\rho)}=\frac{2q^2}{{\cal L}_F\Sigma^2} .
                                                                         \label{VVIP}
  \eeq
For the regular solutions on the interior disk geometry requires
$p_{\perp}+\rho=0$. It follows ${\cal
L}_F\Sigma^2\rightarrow\infty$ and hence ${\cal
L}_F\rightarrow\infty$, since $\Sigma\rightarrow 0$ on the disk.
The case {\bf (B)} represents thus realization of the underlying
hypothesis of non-linearity replacing a singularity.

Third possibility is vanishing of the expression in the square
brackets in the right hand sides of equations (\ref{f3B})-(\ref{f4B})
      \beq
F_{31}\frac{\partial F}{\partial\theta}+F_{23}\frac{\partial
F}{\partial r}=0
                                                                                 \label{f7}
\eeq
 \beq
 F_{01}\frac{\partial F}{\partial\theta} +F_{20}\frac{\partial
 F}{\partial r}=0 .
                                                                                   \label{f8}
  \eeq
With taking into account (\ref{FieldComp}), this system reduces to
    \beq
    a^2\sin^2\theta F_{01}\frac{\partial F}{\partial\theta}
    +(r^2+a^2)F_{20}\frac{\partial F}{\partial r}=0
                                                                                 \label{f9}
  \eeq
  \beq
  F_{01}\frac{\partial F}{\partial\theta}+F_{20}\frac{\partial
  F}{\partial r}=0 .
                                                                                   \label{f10}
 \eeq
It is the system of two algebraic equations for $F_{01}({\partial
F}/{\partial\theta})$ and $F_{20}({\partial F}/{\partial r})$. Its
determinant is equal $Det = -\Sigma$. The case $\Sigma\neq 0$
corresponds to the trivial solutions $F_{01}({\partial
F}/{\partial\theta})=0$, $F_{20}({\partial F}/{\partial r})=0$,
and includes the case

{\bf (C)} $F_{10}=F_{20}=0$, zero fields regime.

\subsection{Necessary condition for the existence of solutions}

The basic question in the case of the system of four equations
(\ref{f1})-(\ref{f4}) for three functions $F_{10}, F_{20}$ and
${\cal L}_F$, is the question of its compatibility, i.e.
compatibility of equations (\ref{DynEq})-(\ref{BianEq}).

\vskip0.1in

The system (\ref{f1})-(\ref{f4}) for $F_{10}, F_{20}$ and ${\cal
L}_F$, with taking into account (\ref{FieldComp}), reduces to
  \beq
(r^2+a^2)\frac{\partial F_{10}}{\partial r} + \frac{\partial
F_{20}}{\partial\theta}=A_1F_{10}+A_2F_{20}
                                                                          \label{e1}
        \eeq
        \beq
        \frac{\partial F_{10}}{\partial
        r}+\frac{1}{a^2\sin^2\theta}\frac{\partial
        F_{20}}{\partial\theta}=B_1F_{10}+B_2F_{20}
                                                                         \label{e2}
  \eeq
  \beq
  \sin^2\theta\frac{\partial
  F_{10}}{\partial\theta}-\frac{(r^2+a^2)}{a^2}\frac{\partial
  F_{20}}{\partial r}=C_1F_{10}+C_2F_{20}
                                                                         \label{e3}
\eeq
  \beq
-\frac{\partial F_{10}}{\partial\theta}+\frac{\partial
F_{20}}{\partial r}=0 ,
                                                                       \label{e4}
   \eeq
   where
    \beq
A_1=-2r-(r^2+a^2)\frac{1}{{\cal L}_F}\frac{\partial{\cal
L}_F}{\partial r}; ~~~A_2=-\cot\theta-\frac{1}{{\cal
L}_F}\frac{\partial{\cal L}_F}{\partial\theta}; ~~~
B_1=-\frac{1}{{\cal
  L}_F}\frac{\partial{\cal L}_F}{\partial r};
 \eeq
 \beq
  B_2=\frac{\cos\theta}{a^2\sin^3\theta}
  -\frac{1}{a^2\sin^2\theta}\frac{1}{{\cal L}_F}\frac{\partial{\cal
  L}_F}{\partial\theta};~~~
C_1=-\sin2\theta; ~~~C_2=\frac{2r}{a^2} .
  \eeq

The system (\ref{e1})-(\ref{e4}) can be resolved with respect to
the derivatives of  $F_{10}$ and $F_{20}$. This gives
  \beq
  \frac{\partial F_{10}}{\partial r}={\tilde A}_1F_{10}+{\tilde
  A}_2F_{20}
                                                                           \label{e5}
  \eeq
  \beq
\frac{\partial F_{20}}{\partial\theta}= {\tilde B}_1F_{10}+
{\tilde B}_2F_{20}
                                                                              \label{e6}
   \eeq
   \beq
   \frac{\partial F_{10}}{\partial\theta}=\frac{\partial
   F_{20}}{\partial r}=
{\tilde C}_1F_{10}+ {\tilde C}_2F_{20} ,
                                                                             \label{e7}
  \eeq
where
  \beq
{\tilde A}_1=-\frac{2r}{\Sigma}-\frac{1}{{\cal
  L}_F}\frac{\partial{\cal L}_F}{\partial r};~ ~~
{\tilde A}_2=-\frac{2\cot\theta}{\Sigma}; ~~~ {\tilde
B}_1=\frac{2a^2r\sin^2\theta}{\Sigma}~;
                                                                              \label{e8}
       \eeq
       \beq
 {\tilde B}_2=\frac{1}{\Sigma}\bigg[\cot\theta\bigg(r^2+a^2+a^2\sin^2\theta\bigg)\bigg]-\frac{1}{{\cal
L}_F}\frac{\partial{\cal L}_F}{\partial\theta};~~~
    {\tilde C}_1=\frac{a^2\sin2\theta}{\Sigma};~~~{\tilde C}_2=-\frac{2r}{\Sigma}.
   \eeq

\vskip0.1in

Equality of the mixed second derivatives of $F_{10}$ and $F_{20}$
gives, respectively,
  \beq
   D_1F_{10}+D_2F_{20}=0,  ~~D_3F_{10}+D_4F_{20}=0 ,
                                                                                  \label{e9}
  \eeq
where
  \beq
  D_1=-\frac{\partial}{\partial\theta}\bigg(\frac{1}{{\cal
   L}_F}\frac{\partial{\cal L}_F}{\partial r}\bigg);~~~
D_2=\frac{2}{\Sigma{\cal L}_F}\bigg[r\frac{\partial{\cal
L}_F}{\partial r}+\cot\theta\frac{\partial{\cal
L}_F}{\partial\theta}\bigg];
  \eeq
  \beq
D_3=\frac{2a^2\sin^2\theta}{\Sigma{\cal
L}_F}\bigg[r\frac{\partial{\cal L}_F}{\partial
r}+\cot\theta\frac{\partial{\cal L}_F}{\partial\theta}\bigg];~~~
D_4=\frac{\partial}{\partial r}\bigg(\frac{1}{{\cal
   L}_F}\frac{\partial{\cal L}_F}{\partial\theta}\bigg) .
      \eeq

\vskip0.1in

We obtained the uniform system of two algebraic equations
(\ref{e9}) with respect to field tensions $F_{10}$ and $F_{20}$.
Necessary and sufficient condition of the existence of a
non-trivial solution of this system is vanishing of its
determinant. Hence, the necessary and sufficient condition of
compatibility of equations (\ref{e1})-(\ref{e4}) is
   \beq
D_1D_4-D_2D_3=0 .
                                                                                   \label{Det}
           \eeq
 In the explicit
form  it reads
   \beq
\frac{\partial}{\partial r}\bigg(\frac{1}{L_F} \frac{\partial
L_F}{\partial \theta}\bigg)\frac{\partial}{\partial
\theta}\bigg(\frac{1}{L_F} \frac{\partial L_F}{\partial
r}\bigg)+\frac{4a^2\sin^2(\theta)}{\Sigma^2}\frac{1}{L^2_F}\bigg[r\frac{\partial
L_{F}}{\partial r}+\cot(\theta)\frac{\partial L_{F}}{\partial
\theta}\bigg]^2=0 .
                                                                                     \label{cond1}
   \eeq
This is the condition on a function ${\cal L}_F$, which is the
necessary and sufficient condition of compatibility of equations
(\ref{DynEq})-(\ref{BianEq}) and hence necessary condition for the
existence of solutions.

 The condition (\ref{cond1}) is evidently satisfied for ${\cal
L}_F=$const which can be normalized to ${\cal L}_F=1$ and
corresponds to the Maxwell weak field limit, and in the case of
trivial zero fields solutions.

The equation (\ref{cond1}) can be written as
   \beq
   \frac{1}{{\cal L}_F^2}\bigg(\frac{\partial^2{\cal L}_F}{\partial
   r\partial\theta}\bigg)^2-\frac{{\cal L}_{FF}}{{\cal L}_F^3}\bigg(\frac{\partial^2{\cal L}_F}{\partial
   r\partial\theta}\bigg)^2\bigg[\frac{\partial
   F}{\partial\theta}\frac{\partial{\cal L}_F}{\partial r} + \frac{\partial
   F}{\partial r}\frac{\partial{\cal L}_F}{\partial\theta}\bigg]+\frac{4a^2\sin^2(\theta)}{\Sigma^2}\frac{1}{L^2_F}
   \bigg[r\frac{\partial L_{F}}{\partial r}+\cot(\theta)\frac{\partial L_{F}}{\partial
\theta}\bigg]^2=0 .
                                                                                 \label{cond2}
    \eeq
   This condition should be analyzed carefully since it implies
   that ${\cal L}_F(r,\theta) \subset C^2$. However, it could be not
   the case on the whole manifold. For regular solutions,
   invariant $F$ vanishes on the disk where it is given by  \cite{me2006}
      \beq
F=-\frac{\kappa^2(p_{\perp}+\rho)^2\Sigma^2}{2q^2} ,
                                                                                \label{FVIP}
  \eeq
and vanishes at infinity in the Maxwell weak field limit.
Lagrangian ${\cal L}(F)$ is a function of a non-monotonic function
$F$ with equal limiting values and should suffer
branching and have a cusp at a certain value of $F$. Correct
description of Lagrange dynamics requires non-uniform variational
problem \cite{us2015}. On the boundary hypersurface which divides
two regions of the manifold with different Lagrangians, the
function ${\cal L}_F$ does not belong to the class $C^2$ as a
function of $F$, but can be $C^2$ as a function of $r$ and
$\theta$. This point has been studied separately and the results
will be reported somewhere \cite{us2016}. Here we point out that
${\cal L}_F$ is finite in the boundary region and is $C^2$ in the
region described by the internal Lagrangian including the interior
region where ${\cal L}_F\rightarrow\infty$. In this region
asymptotic solutions (\ref{f5})-(\ref{f6}) satisfy dynamical
equations (\ref{DynEq})-(\ref{BianEq}), regularity requires ${\cal
L}_F\Sigma^2\rightarrow\infty$ and ${\cal L}_F\rightarrow\infty$,
so that the condition (\ref{cond2}) is satisfied and gives formal
confirmation of compatibility of the system
(\ref{DynEq})-(\ref{BianEq}) in this limit.

\section{Vacuum interiors}

The relation connecting density and pressure with the
electromagnetic fields   reads \cite{me2006}
  \beq
   \kappa (p_{\perp}+\rho)=2{\cal L}_F\left(
F_{10}^2+\frac{F_{20}^2}{a^2\sin^2\theta}\right) .
                                                                           \label{VIPF}
  \eeq
Vacuum ${\cal E}$-surface is defined by $p_{\perp}+\rho=0$.
 By virtue of (\ref{VVIP}) it
leads to ${\cal L}_F\rightarrow\infty$. As a result
 the magnetic permeability vanishes and electric
permeability goes to infinity, so that the ${\cal E}$-surface and
the disk display the properties of a perfect conductor and ideal
diamagnetic. In the limit ${\cal L}_F\rightarrow\infty$ the
magnetic induction vanishes on the vacuum ${\cal E}$-surface and
on the disk by virtue of the asymptotic solutions (\ref{f6}) which satisfy the dynamical equations
(\ref{DynEq})-(\ref{BianEq}) in this limit.

On the de Sitter disk $r=0$ we obtain from (\ref{s2})-(\ref{s4})
$\epsilon_r^r=\epsilon_{\theta}^{\theta}={\cal L}_F$ and
 $\mu_r^r=\mu_{\theta}^{\theta}={{\cal L}_F}^{-1}=\mu\rightarrow 0$.
The magnetic induction $\mathbf{B}$ also vanishes on the disk. In
electrodynamics of continued media the transition to a
superconducting state corresponds to the limits
${\mathbf{B}}\rightarrow 0$ and $\mu\rightarrow 0$ in a surface
current  ${\mathbf{j_s}}=
\frac{(1-\mu)}{4\pi\mu}[{\mathbf{n}}{\mathbf{B}}]$, where
${\mathbf{n}}$ is the normal to the surface. The right-hand side
then becomes indeterminate, and there is no condition which would
restrict the possible values of the current \cite{landau2}. On the
de Sitter disk we can apply definition of a surface current for a
charged surface layer, $4\pi
j_k=[e_{(k)}^{\alpha}F_{\alpha\beta}n^{\beta}]$ \cite{werner},
where $[..]$ denotes a jump across the layer; $e_{(k)}^{\alpha}$
are the tangential base vectors associated with the intrinsic
coordinates on the disk  $t,\phi$, $0\leq\xi\leq\pi/2$;
$n_{\alpha}=(1+q^2/a^2)^{-1/2}\cos\xi~\delta^1_{\alpha}$ is the
unit normal directed upwards \cite{werner}. With using asymptotic
solutions (\ref{f6}) and magnetic permeability $\mu=1/{\cal L}_F$,
we obtain the surface current \cite{portrait}
   \beq
   j_{\phi}=-\frac{q}{2\pi a}
   ~\sqrt{1+q^2/a^2}~\sin^2\xi~\frac{\mu}{\cos^3\xi} .
                                                                                      \label{current}
 \eeq
At approaching the ring $r=0,~\xi=\pi/2$, both terms in the second
fraction go to zero quite independently. As a result the surface
currents on the ring can be any and amount to a non-zero total
value. Superconducting currents flowing (forever)
 on the de Sitter vacuum ring can be considered as a source
 of the Kerr-Newman fields. This kind of a source is non-dissipative
 so that life time of the electromagnetic spinning structure can be unlimited.

 We find the existence of the interior de Sitter vacuum ${\cal
E}$-surface, which contains de Sitter disk as the bridge, with zero magnetic
induction on the whole surface. The next question - what is
going on within ${\cal E}$-surface, in cavities between its upper
and down boundaries and the bridge? Negative value of
$(p_{\perp}+\rho)$ in (\ref{VIPF})~
 would mean  negative values for the electric and magnetic
 permeabilities inadmissible in electrodynamics of continued media
 \cite{landau2}.

 One possibility to satisfy the basic requirement of electrodynamics of continued media,
 can be zero value of $(p_{\perp}+\rho)$ also inside ${\cal E}$-surface.
 This can be the case for the shell-like models
 (\cite{lopez} and references therein) with the flat vacuum
 interior, zero fields and in consequence zero density and
 pressures, and no violation of the weak energy condition.

 The other possibility, favored by the underlying idea of nonlinearity replacing
 a singularity and suggested by vanishing of magnetic induction on the surrounding
${\cal E}$-surface, is  the extension of ${\cal
L}_F\rightarrow\infty$ to its interiors. Then we have de Sitter
vacuum core, $p=-\rho$,
 with the properties of a perfect conductor and ideal diamagnetic,
 zero magnetic induction, and valid
 weak energy condition throughout the whole core.

\section{Summary and discussion}

We studied regular rotating electrically charged rotating black
holes and solitons in nonlinear electrodynamics non-minimally
coupled to gravity with the model-independent approach based on
generic information following from the source-free dynamical
equations, in which nonlinear electromagnetic fields provide a
source in the Einstein equations.

Regular spherically symmetric NED-GR solutions are asymptotically
de Sitter for $r\rightarrow 0$ and asymptotically
Reissner-Nordstr\"om for $r\rightarrow\infty$. They always satisfy
the weak energy condition since $\kappa(p_{\perp}+\rho)=-{\cal
L}_FF$, the invariant $F$ is generically non-positive and ${\cal
L}_F$ represents the electric permeability which can not be
negative in electrodynamics of continued media.

The spherical solutions give rise, by the G\"urses-G\"ursey
algorithm, to regular axially symmetric solutions describing
rotating electrically charged black holes and solitons,
asymptotically Kerr-Newman for a distant observer. Black holes
have at most two horizons and ergoregions. Solitons can have two
ergoregions.

We formulated the necessary and sufficient conditions of
compatibility of dynamical equations governing the electromagnetic
fields dynamics, necessary for the existence of solutions, and
found asymptotic solutions needed to study field dynamics in the
interior regions.

Rotation transforms de Sitter center into de Sitter disk $r=0$
with $p_{\perp}+\rho=0$ and with superconducting current flowing
on the surrounding ring (in place of the Kerr singularity).
Superconducting current can be regarded as a non-dissipative
source of the Kerr-Newman exterior fields, providing unlimited
life time of NED-GR spinning regular objects \cite{portrait}.

 The weak energy condition could be violated in the case when the related
spherical solution satisfies the dominant energy condition. In
this case there exists the vacuum ${\cal E}$-surface defined by
$p_{\perp}+\rho=0$ with the de Sitter disk $r=0$ as a bridge. Both
${\cal E}$-surface and the disk have properties of the perfect
conductor and ideal diamagnetic. Violation of the weak energy
condition inside the vacuum surface would need negative values of
electric and magnetic permeability inadmissible in electrodynamics
of continued media. We can conclude that WEC is not violated in the interiors of
regular rotating charged black holes and solitons in NED-GR theories
compatible with the basic requirement of electrodynamics of continued
media on the
electric and magnetic permeability.

Alternative favored by the underlying idea of non-linearity
regularizing a singularity and suggested by vanishing magnetic
induction on the vacuum ${\cal E}$-surface, is extension of its
basic properties to its interior. Then the regular interior is
presented by de Sitter vacuum core with properties of perfect
conductor and ideal diamagnetic.

\vskip0.1in

This work was motivated by search for an image of the electron,
inspired by the Dirac paper on the extended electron \cite{dirac},
and by the Carter remarkable discovery of ability of the
Kerr-Newman solution to represent the
 electron for a distant observer. Electromagnetic spinning soliton can be
considered as the Coleman lump,  a non-singular, non-dissipative
solution of finite energy holding itself together by its own
self-interaction \cite{coleman}, which can be applied as the model
for the extended electron.

For the electron $q=-e$ and $a=\lambda/2$ where $\lambda=\hbar/m$
is its Compton wavelength \cite{carter}. In the observer region
$r\gg\lambda_e$
   \beq
   E_r=-\frac{e}{r^2}\left(1-\frac{\hbar^2}{m_e^2c^2}\frac{3\cos^2\theta}{4r^2}\right);~
   E_{\theta}=\frac{e\hbar^2}{m_e^2c^2}\frac{\sin 2\theta}{4r^3};
   ~~
   B^r=-\frac{e\hbar}{m_ec}\frac{\cos\theta}{r^3}=2\mu_e\frac{\cos\theta}{r^3};
   ~ B_{\theta}=-\mu_e\frac{\sin\theta}{r^4} .
                                                                                \label{lump}
 \eeq
The Planck constant appears due to discovered by Carter ability of
the Kerr-Newman solution to present the electron as seen by a
distant observer. In terms of the Coleman lump (\ref{lump})
describes the following situation: The leading term in $E_r$ gives
the Coulomb law as the classical limit $\hbar=0$, and the higher
terms represent the quantum corrections.

Purely electromagnetic reaction of annihilation
$e^{+}e^{-}\rightarrow\gamma\gamma(\gamma)$ reveals, at the
$5\sigma$ confidence level, the existence of the minimal length
$l_e=1.57\times 10^{-17}$ cm at the scale $E=1.253$ TeV  which can
be explained as the distance of the closest approach of
annihilating particles where electromagnetic attraction is stopped
by the gravitational repulsion of the interior de Sitter vacuum of
electromagnetic spinning soliton \cite{me2014}.

\end{document}